\documentstyle[12pt,aasms4]{article}

\slugcomment{ to appear in Ap. J. , xx.}

\newcommand{\beq}{\begin{equation}}
\newcommand{\eeq}{\end{equation}}
\newcommand{\etal}{et~al.}
\lefthead{McArthur \etal}
\righthead{Parallax of TV Columbae}

\begin{document}

\title{Interferometric Astrometry with {\it Hubble Space Telescope} Fine Guidance Sensor 3: 
\\ The Parallax of the Cataclysmic Variable \\TV
Columbae} 

\author{B. E. McArthur and G. F. Benedict}
\affil{McDonald Observatory, University of Texas, Austin, Texas 78712}
\author{J. Lee and  W. F. van Altena }
\affil{Department of Astronomy, Yale University, New Haven, Connecticut 05620 }
\author{C. L. Slesnick, R. Rhee, R. J. Patterson and L. W. Fredrick}
\affil{Department of Astronomy, University of Virginia, Charlottesville, Virginia 22903}
\author{T. E. Harrison}
\affil{Department of Astronomy, New Mexico State University, Las Cruces, New Mexico 88003}
\author{W. J. Spiesman}
\affil{McDonald Observatory, University of Texas, Austin, Texas 78712}
\author{E. Nelan}
\affil{Space Telescope Science Institute, Baltimore, MD 21218}
\author{ R. L. Duncombe, P. D. Hemenway \altaffilmark{2}, W. H. Jefferys, and  P. J. Shelus}
\affil{Department of Astronomy, University of Texas, Austin, Texas 78712}
\author{O. G. Franz and L. H. Wasserman}
\affil{Lowell Observatory, Flagstaff, Arizona 86001}
\altaffiltext{2} {Now at Department of Oceanography, University of Rhode Island}



\begin{abstract}
TV Columbae  (TV Col)  is a 13th magnitude  Intermediate Polar (IP) 
Cataclysmic Variable (CV), with multiple periods found in the light
curves.
Past estimates predicted a distance of 400 parsec to greater than 500 parsec. 
Recently completed Hubble Space Telescope (HST)  Fine Guidance
Sensor (FGS) interferometric  observations  allow us to determine the first
trigonometric parallax  to TV Col. This determination puts the distance
of TV Col at $368^{-15}_{+17}$ parsecs.  

CD-32 2376, a 10th magnitude Tycho Catalog star,  is a reference
star in the TV Col frame.  We find a  distance of  $127.7^{-1}_{+1}$ parsecs.

\end{abstract}




%

\section{Introduction}

Cataclysmic variables are binary systems  consisting of a white dwarf (the primary)
that receives mass from the Roche lobe of a  low-mass, late-type
companion (the secondary). 
If the
white dwarf  is strongly magnetized, the binary is  categorized as a
Magnetic Cataclysmic Variable (MCV).
MCVs  have two sub-classes depending upon the magnetic field strength: the Intermediate 
Polars(IP) (sometimes called DQ Hers)  and the Polars (sometimes called AM 
Hers).  
In IPs, the transferred matter from the secondary moves through 
an accretion disk until the magnetic
field  of the white dwarf disrupts the disk and channels the flow
along field lines onto the primary.
X-rays are produced by the shock-heated gas near the surface of the white dwarf,
while the UV and optical emission is mostly from the accretion disk.  
IPs rotate asynchronously, unlike Polars, due to their greater accretion
rate and greater separation (\cite{Pat94}), weaker magnetic fields on the 
primaries (\cite{War95}), and the geometry of the magnetic field (Robinson,
E. L, 2000, personal communication).
IPs represent about 5-10\% of all  CVs, see Patterson(1994) and Warner(1995) 
for  reviews of this class.
Like most other CVs, distances are very uncertain
for most of the IPs (\cite{Berr87}).

TV Col, an IP star,
was first discovered as the hard X-ray source 2A 0526-328
by Cooke \etal (1978).  The X-ray source was optically identifed with 
TV Col by Charles et 
al. (1979), being the first CV discovered by its X-ray emission.
It has an orbital binary period of 5.486 hours (\cite{Hut81}) 
detected from the emission-line radial velocites. 
TV Col shows  four additional  periods:
a 1911  s X-ray period representing the rotation period of 
the white dwarf - the spin period (\cite{Schr85}), 
a 4 day nodal precession of the 
accretion disc period (\cite{Hel93}),  a $\sim 5.2$ hour period 
which is the beat
between the two longer periods ({\cite{Motch81}) - a 
negative superhump (\cite{Rett00}),
and a    photometric period  of 6.36 hours  ({\cite{Rett00}) - a
positive superhump.
It has the 
longest orbital period of any  permanent superhump system.  	
TV Col has 0.1 mag rapid flickering (\cite{Barr88})
and frequent, small-amplitude outbursts of 
luminosity (\cite{Szk84}, \cite{Cord95}). TV Col also has had  
dwarf novae-like short outbursts of 4 magnitude amplitude,  
observed optically
and with IUE 
(\cite{HelB93}; \cite{Hel93};  \cite{Szk84}; \cite{Schw88}; and \cite{Schw87}).

\section{Observations and Reductions}

The observations of TV Col (ICRS 2000: 05 29 25.57, -32 49 95.2)
 were  made with Fine Guidance 
Sensor 3 (FGS3) on the HST.
Astrometry with the HST Fine Guidance
Sensors has been previously described (\cite{Ben94};\cite{Ben93}),
as has the FGS instrument (\cite{Bra91}).
Ten   observations (one orbit each) of TV Col near maximum
parallax factors were made between 1995 and 
1998 with FGS3 in POS (fringe tracking) mode. 
HST FGS parallax observing strategies and reduction and analysis
techniques  have been
described  by  McArthur \etal \space (1999), Benedict \etal \space (1999), 
Harrison \etal  \space (1999), and van Altena \etal \space (1997).

FGS astrometry is relative to a local reference frame. To obtain an absolute parallax for our target requires estimates of the absolute parallaxes of the stars comprising our local reference frame. To obtain these estimates we require
photometry and classification spectra. We obtain JHK photometry from
the second incremental release of the Two Micron All Sky
Survey (2MASS) catalog; B, V, and I from CCD observatiions (obtained at New Mexico State University); and Washington-DDO photometry from  the University of Virginia. We obtain an upper
limit on interstellar absorption in the direction of TV Col from the NASA/IPAC Extragalactic Database (NED)
compilation of the \cite{Sch98} reddening estimates.
This provides
a color excesses, E($\it B-V$), and through the standard relationship,
A$_V$/E($\it B-V$)=3.1, an absorption value, A$_V$.
The effects on the JHK colors are at or below 0.02 magnitude.

We obtained stellar classifications from two sources; the
WIYN telescope\footnote{The WIYN Observatory is a joint 
facility of the University of
Wisconsin-Madison, Indiana University, Yale University, and the National
Optical Astronomy Observatories.}
~multiobject spectrograph (MOS/Hydra) and the CTIO 1.5m with Cassegrain spectrograph. The two independent estimates of spectral 
type and luminosity class are listed in Table 1.

The final adopted spectral types are the result of plotting the 
photometry (collected in Table 2) on several color-color 
diagrams ($\it B-V$ vs $\it V-K$ and $\it J-K$ vs $\it V-K$) 
upon which are impressed a 
mapping between colors and spectral types
from Bessel \& Brett (1988) and Allen's Astrophysical 
Quantities 4$^{th}$ edition (\cite{Cox00}, hereafter AQ2000).  Because 
both sources
for our spectral types and luminosity classes
 expressed some doubt as to the luminosity class, 
we use Washington-DDO photometry (\cite{Maj00}) to confirm 
their estimates. Plotting our reference stars on the giant/dwarf 
discrimination plane (M-D vs M-T$_2$, Figure~\ref{fig-1}), reference 
stars 1, 2, and 
3 are clearly dwarfs. Ref-4 is borderline dwarf/giant or a metal poor dwarf. 
Because we obtain a  better solution (the ratio of $\chi^2$ to the number of degrees of freedom is smaller), we adopt a dwarf luminosity class for 
ref-4. V magnitude and colors are listed in Table~\ref{tbl-2}.

Absolute magnitudes,
M$_V$, are taken from the AQ2000 tables as a function
of spectral type. We assume an error for each M$_V$ consistent
with the spectral type and luminosity class differences among WIYN, 
NMSU, and Washington-DDO photometry (Tables~\ref{tbl-1},\ref{tbl-2},
\ref{tbl-3}).
The resulting absorption-corrected distance moduli with errors and derived
parallaxes are presented in Table~\ref{tbl-3}.
These are the parallaxes and associated errors used in the modelling.
The spectrophotometric parallax of ref-1 and our derived value agree
within the errors, providing confidence in our approach.

The average color of
the reference stars and our science target differ, with $\Delta(\bv) \sim -0.91$.
Therefore,  we apply the differential correction for lateral color discussed in
\cite{Ben99} 
to the TV Col  observations. The  GaussFit solved equation
of condition for TV Col becomes:
\beq
 	x' = x + lc(\it B-V) 
\eeq
\beq
 	y' = y + lc(\it B-V) 
\eeq
\beq
\xi = A*x' + B*y' + C + R_x \times (x'^2 + y'^2) - \mu_x  - P_\alpha\pi_x
\eeq
\beq
\eta = -B*x' + A*y' + F + R_y \times (x'^2 + y'^2) - \mu_y  - P_\delta\pi_y
\eeq
where $\it x$ and $\it y$ are the rectangular HST coordinates, 
$\it lc$ is the derived
lateral color correction (\cite{Ben99}); $\it B-V $ is 
the  B-V  magnitude; A  and  B   
are a set of scale plate constants, C and F are
zero point corrections, $R_x$ and $R_y$ are radial terms, 
$\mu_x$ and $\mu_y$ are proper motions,  
$P_\alpha$ and $P_\delta$ are parallax factors,  and $\it \pi_x$ and $\it \pi_y$ are  the parallaxes in x and y.
The spectrophotometrically determined parallaxes of the reference
frame stars are modelled simultaneously as observations with errors 
to produce an absolute, not relative parallax for TV Col.

Comparing Tables~\ref{tbl-3} and ~\ref{tbl-4} we find that the final adjusted
absolute parallaxes for the reference stars agree with the spectrophotometric
estimates within the errors.  Our derived absolute parallax for TV Col is
given in Table~\ref{tbl-5}, along with a relative proper motion.
As seen in Table~\ref{tbl-4},
the standard errors  resulting from the solutions of the equations for
parallax and proper motion
are sub-milliarcsecond.   Figure~\ref{fig-2}
shows histograms of the the residuals from the fit of the target and
the reference frame stars. 
The histogram of residuals is by far the best we have seen in dealing
with over twenty FGS parallax data sets.
Typical Gaussians fits are characterized by
sigmas of order 1 mas. This was obviously an
astrometrically very quiet reference frame.  
A generous well-characterized reference frame  surrounding the
target along with  exceptional instrument performance evidenced
by the small standard error of the guider FGS's contributed to
the very low internal errors for these observations.
Typical internal errors for FGS parallaxes range 0.3 - 0.7 mas.
This 0.1 mas internal error is the luck of the draw. Our systematic
errors in parallax determinations are likely larger.




\section{Trigonometric Parallax and Absolute Magnitude of TV Col}

Using the weighted spectrophotometric parallaxes of the reference
frame, the simultaneous modelling of the observations gives an HST  parallax for
TV Col of 2.7{$\pm$}0.11 mas,  
for a distance of   $368^{-15}_{+17}$ parsecs 
(Table~\ref{tbl-5}). The distance  
estimates from  other non-astrometric methods are listed in Table~\ref{tbl-6}.

The distance modulus $(- 5 + 5  \log (1/\pi))$ for TV Col is 7.84.
Using Bruch and Engel's (1994) visual magnitude 
of 13.75 $\pm 0.1$  for the apparent magnitude
we obtain an absolute magnitude ($M_V$) of $5.92^{-0.1}_{+0.1}$.
Adjusting the distance modulus with a correction for $A_{v}$
$(- 5 + 5  \log (1/\pi)-A_{v})$  give an absolute magnitude of $5.81^{-0.1}_{+0.1}$.
Ritter and Kolb (1998) list visual magnitudes of 13.6 and 14.1.
When using a trigonometric parallax to estimate the absolute
magnitude of a star, a correction should be made for the
Lutz-Kelker (LK) bias (\cite{Lutz73}).
Because of the galactic latitude and distance of TV Col, 
and the scale height of the
stellar population of which it is a member,
we do not use a uniform space density  for calculating
the LK bias, but use a density law that falls off as the 
-0.5 power of the distance at the distance of TV Col.
This translates into $\it n $ = -3.5 as the power in the parallax distribution.
This $\it n$ is then used in an LK
algorithm modified by Hanson (H)(1979) to include the power law
of the
parent population. A correction of -0.03 $\pm$ 0.01 mag is derived for
the
LKH bias, which makes the absolute magnitude $5.89^{-0.12}_{+0.12}$
$(5.78^{-0.12}_{+0.12}$ with the $A_{v}$ correction).

TV Col (Galactic coordinates: {\it l} = 236.79, {\it b} = -30.60) is 
well below the plane of the disk of the  galaxy. 
From our absolute parallax  and relative proper motion we derive a 
space velocity of 103 km s$^{-1}$ (relative to our reference frame).
The small difference between the relative proper motion derived for
CD-32 2376 (50.097  mas y$^{-1}$) and 
the absolute motions listed in the USNO ACT (50.604  mas y$^{-1}$) and
Tycho-2 (49.519  mas y$^{-1}$) catalogs, see Table~\ref{tbl-7}, suggests a   
small difference between relative and absolute proper motion.
The velocity component perpendicular to the galactic plane, $\bf W$,  is -5.1
km s$^{-1}$.
Our new parallax places the star 187 parsecs 
below the sun, which locates it 195 parsecs below the galactic
plane. Discussion of space velocity in cataclysmic variables can be
reviewed in these papers: \cite{Spro96}, \cite{Stehle}, and \cite{Par96}.

 \section{The Mass of TV Col - an Unresolved Issue}

ROSAT observations of TV Col (\cite{Vrtil96}) indicated a ratio of X-ray
luminosity to UV and optical luminosity of ~0.2 in quiescence.
Cropper \etal (1999) analyzed  the continuum spectra obtained by Ginga  
to estimate a white dwarf
mass of  1.2 ${\cal M}_{\sun}$,
while  Ishida \& Ezuka (1999) used $\it ASCA$ data to derive a mass 
of 0.51 $^{-0.22}_{+0.41}$ ${\cal M}_{\sun}$.
Ramsay (2000) used $\it RXTE$ data to determine a mass of 
0.96 $^{-0.5}_{+0.2}$ ${\cal M}_{\sun}$.
Radial velocity measurements by Hellier  (1993) provided a $\it K_1 $
of $153\pm 12$
{km s$^{-1}$} corresponding to a mass function of $0.085 \pm 0.020$ ${\cal M}_{\sun}$.
The inclination is estimated to be $70^\circ \pm 3^\circ$ by the width
and depth of the eclipse (\cite{Hell91}).  Assuming that it is a ZAMS star, 
Hellier (1993)
estimated a secondary mass of
0.56 ${\cal M}_{\sun}$ and a primary mass of $0.75 \pm 0.15$ ${\cal M}_{\sun}$ 
(using the  Patterson (1984) equation).
The secondary star is most likely a late K or early M
(\cite{Beuer00}, \cite{Smit98}),
although it is reported as a K1 in SIMBAD, and a K1-5 in Ritter and Kolb (1998).

A measurement of the total separation of the
components, combined with a parallax,  would provide a direct 
determination of their masses. The total estimated mass and measured 
period implies  a = 0.0080AU. Our parallax would set the total 
separation at 21 microarcsecond ($\mu$arcsec) with individual orbits 
of 9 and 12 $\mu$arcsec. TV Col is a potential target for the Space 
Interferometry Mission (http://sim.jpl.nasa.gov). The component 
orbits  are larger than the anticipated (1-2 $\mu$arcsec) SIM narrow 
angle astrometric measurement limits. With $\Delta V \sim 4$, much of 
the system flux is contributed by the WD. Longward of 700 nm the M 
dwarf - WD magnitude difference should decrease somewhat. The SIM 
sensitivity (V$_{lim}\sim20$), wide bandpass (400-1000nm), and 
spectral resolution (R=80) should allow measurement of positions, 
magnitudes, and colors for both components. To derive a precise 
separation will require 5-10 such measurements to establish the 
component A and B orbits and the  mass fraction. SIM 
could also provide an absolute parallax two orders of magnitude more 
precise than that reported here. Together orbits and parallax could 
provide masses with $\sim5$\% error.

\section{Trigonometric Parallax and Absolute Magnitude of CD-32 2376}

One of the reference stars in our field was a 10.46 magnitude
Tycho Catalog Proper
motion star.   The HST parallax of CD-32 2376 was 7.83 $\pm$ 0.08 mas,
which puts the distance at $127.7 ^{-1}_{+1}$. 
We derived a relative $\mu_\alpha$ = 36.9 $\pm$ 0.1 mas y$^{-1}$  and 
relative $\mu_\delta$ = 33.9 $\pm$ 0.2 mas y$^{-1}$
for this star. Table~\ref{tbl-7} compares the HST proper motions with those
listed in the 
Tycho 2 catalog (\cite{Hog00}) and
the ACT (\cite{Urb98}). We attribute the difference in proper motion 
position angle to our very sparse and local reference frame.


%
%

\section{Summary}

HST trigonometric parallaxes can
provide accurate distances  to CVs.  Accurate distances allow  basic quantities
such as absolute magnitude and mass transfer rate to be derived.
Understanding the physics of CVs depends upon these derivations.

\acknowledgments

This research has made use of NASA's Astrophysics Data
System Abstract Service and the SIMBAD Stellar Database inquiry and retrieval
system. 
This research has made use of the NASA/IPAC Extragalactic Database (NED)
which is operated by the Jet Propulsion Laboratory, California 
Institute of
Technology, under contract with the National Aeronautics and Space
Administration. 
This publication makes use of data products from the Two Micron All Sky
Survey, which is a joint project of the University of Massachusetts and the
Infrared Processing and Analysis Center/California Institute of Technology,
funded by the National Aeronautics and Space Administration and the National
Science Foundation.
This work is based on observation made with the NASA/ESA Hubble Space
Telescope, which is operated by  the Space Telescope Science Institute,
of  the Association of Universities for Research in Astronomy, Inc.,
under NASA contract NAS5-26555.
The HST Astrometry Science Team receives support through NASA grant
NAG5-1603.
We thank  Bill Welsh and Rob Robinson
for
helpful discussions and draft paper reviews.
Denise Taylor and Lauretta Nagel provided 
assistance at the Space Telescope Science Institute.

\appendix

\clearpage

\begin{center}
\begin{deluxetable}{cccc}
\footnotesize
\tablecaption{Reference Star Spectral Types\label{tbl-1}}
\tablewidth{0pt}
\tablehead{
\colhead{Star} &
\colhead{NMSU} &
\colhead{WIYN}   &
\colhead{Adopted Spectral Type} 
}
\startdata
 Ref-1 &  G2 V &   G2 IV &   G2 V \nl
 Ref-2 &  G5 V & G1 V &   G2 V \nl
 Ref-3 &  K1/2 V&    K2 V &   K2 V  \nl
 Ref-4 &  K2 V &   G9 V &  G9 V \nl
\enddata
\tablenotetext{}{NMSU = New Mexico State University}
\tablenotetext{}{WIYN  = University of Wisconsin-Madison, Indiana University, 
Yale University, and the National Optical Astronomy Observatories}
\end{deluxetable}
\end{center}

\clearpage
\begin{center}
\begin{deluxetable}{ccccccc}
\footnotesize
\tablecaption{Reference Star Photometery\label{tbl-2}}
\tablewidth{0pt}
\tablehead{
\colhead{Star} &
 \colhead{$\it V$ \tablenotemark{a}} &
 \colhead{$\it B-V$ \tablenotemark{a}}& 
 \colhead{$\it V-K$ \tablenotemark{b}}& 
\colhead{$\it J-K$ \tablenotemark{b}}&
 \colhead{$\it M-D$ \tablenotemark{c}} &
\colhead{$\it M-T_2$ \tablenotemark{c}}   
}
\startdata
 Ref-1& 10.46 &  0.62 $\pm$  0.02&       1.50 $\pm$  0.03 & 0.41 $\pm$  0.04 &  0.01 $\pm$   0.04  &  0.87 $\pm$   0.02 \nl
 Ref-2 & 14.66 &  0.59 $\pm$  0.04 & 1.58 $\pm$  0.04  &  0.34 $\pm$  0.04 & 0.00  $\pm$  0.04 &   0.92  $\pm$  0.02\nl
 Ref-3 & 13.65 &0.96 $\pm$  0.03 &2.20 $\pm$  0.04  & 0.53 $\pm$  0.04 & -0.16  $\pm$  0.04  & 1.18 $\pm$   0.02\nl
 Ref-4 & 13.82 &0.72 $\pm$  0.03 &1.70 $\pm$  0.04 & 0.44 $\pm$  0.04 & -0.01 $\pm$   0.04 &  1.14 $\pm$   0.02\nl
\enddata
\tablenotetext{a}{B, V, and I are from New Mexico State}
\tablenotetext{b}{J, K are from 2MASS}
\tablenotetext{c}{M, D, $T_2$ are Washington-DDO filters}
 
\end{deluxetable}
\end{center}

\clearpage
\begin{center}
\begin{deluxetable}{cccccc}
\footnotesize
\tablecaption{Reference Star Spectroscopic Parallaxes \label{tbl-3}}
\tablewidth{0pt}
\tablehead{
\colhead{Star} &
\colhead{$\it V$} &
 \colhead{$M_V$\tablenotemark{a}} &
\colhead{$A_{v}$\tablenotemark{b}} &
\colhead{m-M-$A_{v}$} &
 \colhead{$\pi_{abs}$}
}
\startdata
Ref-1&     10.46 &   4.7 $\pm$    0.1   &  0.1 &    5.7 & 0.0076$\pm$ 0.0007 \nl
Ref-2 &    14.66 &  4.7  $\pm$  0.1  &   0.1   &  9.9  & 0.0017$\pm$ 0.0004 \nl
Ref-3 &    13.65 &  6.2  $\pm$   0.5  &   0.1  &   7.4  & 0.0022$\pm$ 0.0002 \nl
Ref-4 &    13.82 &  5.7 $\pm$    0.3  &  0.1 &   8.0 &   0.0025$\pm$ 0.0002\nl
\enddata
\tablenotetext{a}{$M_V$ from AQ2000}
\tablenotetext{b}{$A_{V}$ from Schlegel $\etal$  1998}

\end{deluxetable}
\end{center}

\clearpage

\begin{center}
\begin{deluxetable}{crrrrrrrrrrr}
\footnotesize
\tablecaption{TV Col and Reference Star Data \label{tbl-4}}
\tablewidth{0pt}
\tablehead{
\colhead{Star} & 
\colhead{$\xi$  \tablenotemark{a}}   &
 \colhead{$\eta$  \tablenotemark{a}} &
\colhead{$\sigma {_\xi}$ (mas)}   &
 \colhead{$\sigma {_\eta}$ (mas) } &
 \colhead{$\pi_{abs}$ (mas) } 
}
\startdata
TV Col &  31.354 & 667.018 &   0.34 &      0.45 & 2.717  \nl
Ref-1 (CD32-2376)   & 101.968 & 730.702 & 0.30    &   0.37 &7.831  \nl
Ref-2   & 2.500 & 784.643 &  0.44  &   0.72 & 1.681  \nl
Ref-3   & 35.568 & 659.404 & 0.34 &      0.54 & 2.193 \nl
Ref-4   &  -364.898 & 695.884 & 0.4   &      0.5 &2.502 \nl
\enddata
\tablenotetext{a}{$\xi$ and $\eta$ are relative positions in arcseconds }
\end{deluxetable}
\end{center}

\clearpage

\begin{center}
\begin{deluxetable}{lll}
\tablecaption{TV Col Parallax and Proper Motion \label{tbl-5}}
\tablewidth{0in}
\tablehead{\colhead{Parameter} &  \colhead{Value}}
\startdata
{\it HST} study duration  &2.5 y\\
Number of observation sets    &   10 \\
Number of ref. stars &  4  \\
{\it HST}  Parallax & 2.7 $\pm$ 0.11  mas\\
{\it HST}  Distance & $368^{-15}_{+17}$ parsecs\\
{\it HST} Relative Proper Motion  $\mu_\alpha$  & 25.99 $\pm$ 0.1 mas y$^{-1}$\\
{\it HST}  Relative Proper Motion  $\mu_\delta$  & 9.67 $\pm$ 0.2 mas y$^{-1}$\\
{\it HST} Relative Proper Motion ($\mu$)  &27.72 $\pm$ 0.22 mas y$^{-1}$ \\
\indent in p.a. & $69.6^\circ$ \\
\enddata
\tablenotetext{a}{the relative proper motion position angle is likely to vary significantly from absolute because of the sparse reference frame (indicated by p.a. of CD-32 2376)}
\end{deluxetable}
\end{center}
\clearpage

\clearpage

\begin{center}
\begin{deluxetable}{lll}
\tablecaption{CD-32 2376 Parallax and Proper Motion \label{tbl-7}}
\tablewidth{0in}
\tablehead{\colhead{Parameter} &  \colhead{Value}}
\startdata
{\it HST} Parallax & 7.83 $\pm$ 0.08  mas\\
{\it HST} Distance & $127.7^{-1}_{+1}$ parsecs\\
{\it HST} Relative Proper Motion  $\mu_\alpha$  & 36.9 $\pm$ 0.1 mas y$^{-1}$\\
{\it HST}  Relative Proper Motion  $\mu_\delta$  & 33.9 $\pm$ 0.1 mas y$^{-1}$\\
{\it HST} Relative Proper Motion ($\mu$)  &50.097 $\pm$ 0.14 mas y$^{-1}$ \\
\indent in p.a. & $47.4^\circ$ \\
{\it Tycho-2} $\mu$  &49.519 $\pm$ 1.6 mas y$^{-1}$ \\
 \indent in p.a. & $37^\circ$ \\
USNO ACT $\mu$  &50.604 $\pm$ 4.3 mas y$^{-1}$ \\
 \indent in p.a. & $35^\circ$ \\
\enddata
\end{deluxetable}
\end{center}
\clearpage


\clearpage
 
\begin{center}
\begin{deluxetable}{llcc}
\footnotesize
\tablecaption{Distance Estimates to TV Col\label{tbl-6}}
\tablewidth{0pt}
\tablehead{
\colhead{Reference}    &
 \colhead{Distance in parsecs} & \colhead{Method} }
\startdata
Patterson(1984) & 160 & Disk properties \nl
Patterson(1994) & 400 & Photometric parallax of the secondary \nl
Mouchet \etal (1981) & $<500$ & Weak UV interstellar absorption \nl
Bonnet-Bidaud \etal (1985) & 500 & Infrared photometry of secondary\nl
Buckley \& Tuohy (1989)  & $>500$ & Xray distance \nl
{\it HST} (2000) & $368^{-15}_{+17}$  &Trigonometric parallax, this paper\nl
\enddata

\end{deluxetable}
\end{center}

\clearpage

\begin{figure}
\epsscale{1}
\plotone{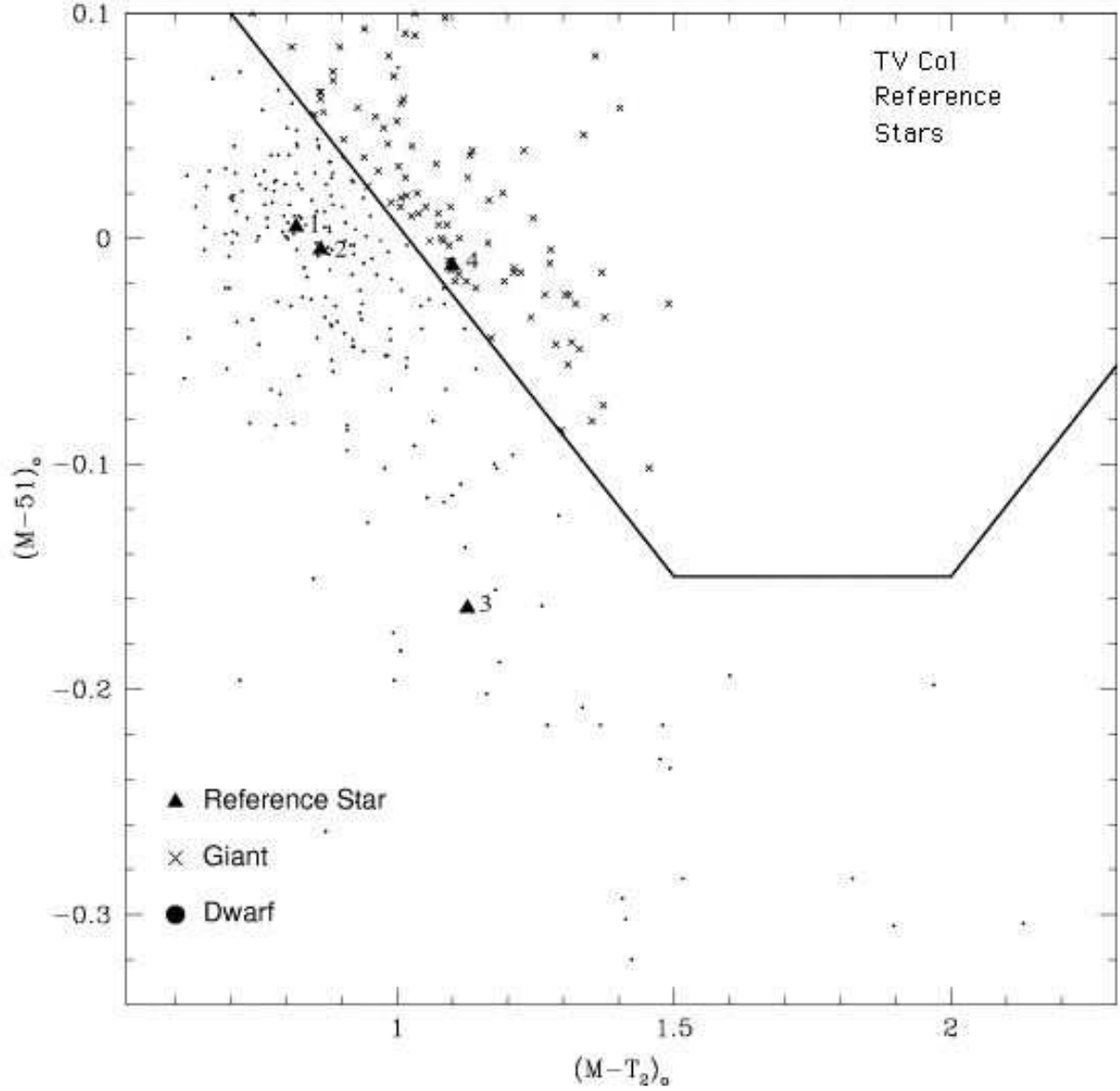}
\caption{Astrometric reference stars plotted on the M-D vs M-T$_2$ plane. The values have been corrected for absorption. Also plotted are known dwarf ( $\cdot$ ) and giant ($\times$) stars. Ref-4 is ambiguous. A dwarf luminosity classification for all reference stars results in a lower $\chi^2$ sor our TV Col parallax solution.} \label{fig-1}
\end{figure}
\clearpage

\begin{figure}
\epsscale{.6}
\plotone{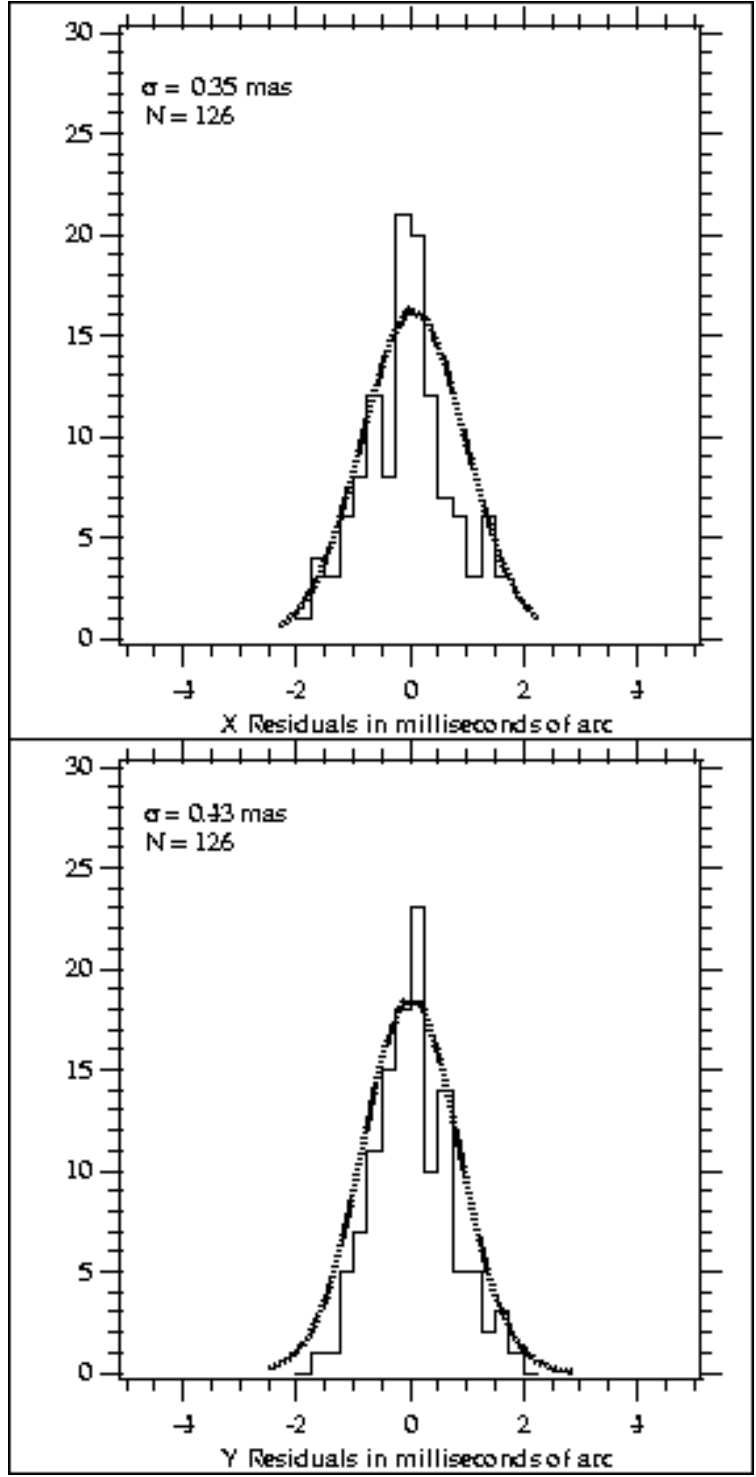}
\caption{Histograms of x and y residuals obtained from modelling  TV Col 
and its reference frame.   Distributions are fit with 
Gaussians.} \label{fig-2}
\end{figure}
\clearpage

\end{document}